\documentclass[runningheads]{llncs}

\usepackage[toc,page]{appendix}

\usepackage[english]{babel} 
\usepackage{ucs}
\usepackage[utf8x]{inputenc}
\usepackage[T1]{fontenc}
\usepackage{pslatex}
\usepackage{graphicx}
\usepackage{xspace}
\usepackage{float}
\usepackage{color}
\usepackage{alltt}
\usepackage{enumitem}
\usepackage[]{algorithm}
\usepackage[noend]{algpseudocode}
\algblock[Name]{Start}{End}
\newcommand{\minarets}{\textsf{Minarets}\xspace}
\renewcommand{\read}{\textbf{Read}\xspace}
\newcommand{\print}{\textbf{Print}\xspace}
\usepackage{setspace}
\usepackage{enumitem}   
\usepackage{fancyvrb}   
\usepackage{tcolorbox}   
\usepackage{verbatimbox}
\usepackage{float}
\usepackage{chngcntr}   
\usepackage{enumitem}
\usepackage{pagecolor}
\tcbuselibrary{skins}

\usepackage{amsmath,bm,times}

\usepackage[pdftex]{thumbpdf} 
\usepackage[bookmarks=true,bookmarksnumbered=true,colorlinks=false,urlcolor=blue,citecolor=blue,linkcolor=blue]{hyperref}
\usepackage{cite}

\def\repHOME{.} 
\def\repBIBLIO{\repHOME/.} 

\title{Multi-Facets Contract for Modeling and Verifying Heterogeneous Systems}
\author{A. Abdelkader Khouass\inst{1,2} \and  J. Christian Attiogbé\inst{1} \and E. Mohamed Messabihi\inst{2}}
\institute{LS2N CNRS UMR 6004 - University of Nantes\\
	\email{\{name.surname\}@univ-nantes.fr} \and
	LRIT - University of Tlemcen\\	
	\email{abderrahmaneabdelkader.khouass@univ-tlemcen.dz }\\ \email{mohamedelhabib.messabihi@univ-tlemcen.dz }}

\begin{document}
	\maketitle
	
	\begin{abstract}
		Critical and cyber-physical systems (CPS) that exist in large industries, such as nuclear power plants, railway, automotive or aeronautical industries are complex heterogeneous systems.
		They are complex because they are open, perimeter-less, often built by assembling various heterogeneous and interacting components which are frequently reconfigured due to requirements.
		Consequently, the modeling and analysis of such systems is a challenge in software engineering.
		We introduce a new method for modeling and verifying heterogeneous systems.
		The method consists in: equipping individual components with \textit{generalized contract}, ordering these contracts according to given \textit{facets}, composing these components and verifying the resulting system with respect to the facets. We illustrate the use of the method by a case study.
		The proposed method may be extended to cover more facets, and by strengthening assistance tool through proactive aspects in modelling and property verification.
	\end{abstract}

\section{Introduction}





Critical and cyber-physical systems (CPS) that exist in large industries, such as nuclear power plants, railway, automotive or aeronautical industries are complex heterogeneous systems. They do not have a precise perimeter, they are open and often built by assembling various components. Their complexity forces one to have a wide variety of heterogeneous components, frequently reconfigurable due to requirements evolution.\\
In addition, the significant competition between suppliers in the last 30 years, has pushed manufacturers to import entire subsystems (components) in order to integrate them into their systems. But, this triggered late errors that impacted the integration process, and caused significant delays and over-costs \cite{Benveniste2018}. The purpose of pre-developed components, with the policy of Components Off The Shelf (COTS)\cite{COTS2020}, is to avoid developing a system from scratch; but, to reuse and adapt these components as needed. This technique reduces time and cost of software development if the integration is well mastered; however, pre-developed components are available often in different languages ​​and in different environments, with their own characteristics.


With the time and advent of concurrent and distributed systems, Component Based Software Engineering (CBSE)\cite{CBSE} has known a high interest. The construction of a distributed system involves several specific components, that respect properties of different types; this requires rigor, methods and tools. Moreover the involved components may deal with various \textit{facets} such as data, interaction, time constraints, security, etc. Therefore, if the integration of components into a global system is not mastered, it generates considerable time losses and overcharges, because of inconsistency of requirements, incompatibility of meaning and properties, late detection of composition errors, etc\\
For all these reasons, the modeling of such global heterogenous systems, is challenging. It is also the case for their formal analysis. The use of efficient methods and techniques is required to meet these challenges.

We aim at studying and alleviating the difficulties of practical modelling and integration of heterogeneous components.  
In this paper, we propose a novel approach based on contracts for modeling and verifying
complex and heterogeneous systems. Our approach (named "ModelINg And veRifying
hEterogenous sysTems with contractS" (\minarets)) consists in modeling and verifying a system with the concept of \textit{generalized contracts}.
The contract is generalized in the
sense that it will allow one to manage the interaction with the components through \textit{given facets}: the
properties of the environment, the properties of the concerned components, the communication
constraints and non-functional properties (quality of service for example).
The use of contracts during the verification reduces the complex manipulation of heterogeneous systems; moreover, the structuring of contracts by facets and priority of properties makes it possible to not only weaken more and more the difficulty of
checking heterogeneous systems, but also, to save time and increase performance during verification.

The rest of the article is structured as follows. Section 2 the preliminaries, Section
3 the modeling and verification methodology, Section 4 illustrates our approach with
experimentation and assessment; Section 5 provides an overview of related work, and
finally, Section 6 presents the conclusion and future work.

\section{Preliminaries}
\label{section:Background}
We present in this section the challenges and features of heterogeneous systems that we used in our method.
\subsection{Challenges and features of heterogeneous systems}
Among the challenges of heterogeneous systems we have the following.
\paragraph{Contract-based assertions}
 were first defined in \cite{Flo1967} and \cite{Hoa1969} where contracts are predicates used as pre-conditions $P$ and post-conditions $Q$ of program statements: $\{P\}~~S~~\{Q\}$. In computer science, a contract is an agreement between a 'server' and a 'client' or between two or more components, in order to ensure a good interaction and finally to satisfy one or more properties. The first component (server) offers a functionality and the other (client) requires a functionality. 

Design by Contract (DBC)\cite{Mey1992} extends the notion of contract to deal with software errors and to increase the reliability of object-oriented systems \cite{Mey1992}. DBC is a method inspired by the legal definition of \textit{contract}; but here, the defined relationship is between a client and a server;  the organization of the structure of contracts between the client and the server is done in the form of (contract) clauses, which are pre-conditions that must be satisfied at the input of a program, post-conditions that must be satisfied at the output of the program and invariants that must be satisfied throughout the execution of the program \cite{Mey1992}. The client that requests an operation from the server must first satisfy the pre-conditions of the operation so that the server can satisfy its post-conditions.

\paragraph{Heterogeneous systems}
A heterogeneous system refers to the composition of a software system with different components; it is composed of modules/components or software that are developed in different languages, in different environments, running in different platforms and with different communication standards, etc. A component may cover a given facet, deal with specific properties or requirements. A functionality offered by such a system is often generated after the interaction between various components.
In a heterogeneous system, several components interact to provide global functionalities; each component may require a functionality to work properly or offer a functionality to another component and vice versa. The components may be either synchronized components that have a global clock and interact on atomic transactions, or asynchronous components which operate with independent clocks and interact in a non-atomic manner. In addition, the interaction method can be a point to point, a diffusion or a group method.
The composition of such various heterogeneous components that must interact with each other to satisfy a given properties is not straightforward; their modeling and verification are a challenge, because of their various features which hamper composition and formal analysis\cite{Att2017,SifBasu2010,Benveniste2007,Benveniste2018,Benveniste2014,Lee2010}. 

\paragraph{Formal verification}
The specification of complex systems in a rigorous manner requires formal methods \cite{Win1990}; these methods, because they are based on mathematical notions help to specify with precision; they allow one to specify a system in a clear and understandable way not only by designers but also by software tools. One advantage of using formal methods is that they provide supports to a \textit{formal verification}, in order to be able to detect errors that are not easily noticeable in running time. These errors can be corrected in the earlier phases of development. However, as heterogeneous systems integrate several components with various properties and facets, their verification still need practical modeling and verifying methods and tools, to alleviate composition difficulties. \\ \\
Some features of heterogeneous system are listed bellow. \paragraph{Composition}
Because a component requires a functionality offered by another one, it can not be directly usable, i.e. the call of one of its functionalities will not work if the components are not bound in advance. The composition of components consists in binding them in an interaction relationship  \cite{Tmessabihi}.  

\paragraph{Composability}
Composability means stability of component properties across integration. That means given components can be composed as they are. A component’s essential properties are not affected during the system construction process, i.e., even after composing the different components together, their essential individual properties are preserved\cite{Basu2010}. 

\paragraph{Compositionality}
The Compositionality denotes the inferring of global system properties from the local properties of the components. An example is inferring global deadlock-freedom from the deadlock freedom of the individual components\cite{Sif2014}.

\begin{definition}[Labelled Transition System LTS]\label{LTS}
	Let $\mathcal{A}$ denote an infinite set of actions, including the invisible action $\tau$ , which denotes internal behaviour. All other actions are called
	visible. An LTS is a tuple ($\sum_{}^{}$, $A$, $\xrightarrow[\text{}]{\text{}}$, $p_{init}$), where $\sum_{}^{}$ is a set of states, A ⊆ $\mathcal{A}$ is a set of actions, $\xrightarrow[\text{}]{\text{}}$ $⊆ $ $\sum_{}^{}$ × A × $\sum_{}^{}$ is the (labelled) transition relation, and
	p$_{init}$ $\in$ $\sum_{}^{}$ is the initial state. We write p
	$\xrightarrow[\text{}]{\text{a}}$ $p'$ if (p, a, $p'$) $\in$ $\xrightarrow[\text{}]{\text{}}$ and p $\xrightarrow[\text{}]{\text{$\tau$*}}$ $p'$ if there is a (possibly empty) sequence of $\tau$ -transitions from p to $p'$, i.e., states p$_{0}$, . . . , p$_{n}$ (n ≥ 0) such that $p = $ $p_{0}$, $p'$ = p$_{n}$, and p$_{i}$ $\xrightarrow[\text{}]{\text{$\tau$}}$ p$_{i+1}$$\ for\ i = 0, . . . , n−1$ \cite{LTS} .	
\end{definition}
\begin{definition}[Parallel composition of LTS]\label{LTSPC} 
	Let $P =$ $($$\sum_{}^{}$$_{P}$ , $A_{P}$$,\ \xrightarrow[\text{}]{\text{}}$$_{P}$,$\ p_{init})$,$\ Q =$ $($$\sum_{}^{}$$_{Q}$,$\ A_{Q}$,\\$\  \xrightarrow[\text{}]{\text{}}$$_{Q}$$,\ q_{init}$$)$, and\ $A_{sync}$$\ ⊆ \mathcal{A}$$\ \textbackslash\  \{\tau\}$. The parallel composition of P and Q with synchronization on $\ A_{sync}$, $"P\ |[A_{sync}]|\ Q"$, is defined as $($$\sum_{}^{}$$_{P}$ × $\sum_{}^{}$$_{Q}$,$\ A_{P}$ ∪ $A_{Q}$, →,$\ (p_{init}$,$\ q_{init}))$, where $(p,\ q)$ $\ \xrightarrow[\text{}]{\text{a}}$ $\ (p',\ q')$ if and only if either (1)$\ p$ $\ \xrightarrow[\text{}]{\text{a}}$ $p'$, $q'$ = $q$, and $a$ $\notin$ $A_{sync}$, or (2) $p'$ = $p$, $q$ $\ \xrightarrow[\text{}]{\text{a}}$ $q'$, and $a$ $\notin$ $A_{sync}$, or (3) $p$ $\ \xrightarrow[\text{}]{\text{a}}$ $p'$, q$\ \xrightarrow[\text{}]{\text{a}}$ $q'$, and $a \in A_{sync}$ \cite{LTS}.
	
\end{definition}

\subsection{Materials}	
In the proposed method, we use an extension of contracts to extend the interface of components; then we consider the composition of components which behaviours are modelled with labelled transition systems. 
\begin{definition}[Contract] \label{contract}
	A contract C is a pair (A, G), where A, the assumption, and G, the guarantee. The assumption must be respected at the entry of the system and the guarantee must be fulfilled at the exit of the system \cite{Mey1992}.   
\end{definition}
\paragraph{Property Specification Language (PSL)}
is a formal specification language \cite{PSL} used to specify the properties and behaviour of systems. It is an extension of the standard temporal logics Linear Temporal Logic (LTL) and the Computation Tree Logic (CTL). PSL has a specification form which could be used as input for formal verification, formal analysis, simulation and hybrid verification tools. PSL improves communication between designers, architects and verification engineers to increase productivity during the design and verification process\cite{PSL}. In our case we use the ALDEC Active-HDL \footnote{https://www.aldec.com/en/products/fpga$\_$simulation/active-hdl} tool that supports PSL. 

\paragraph{UPPAAL}
\cite{uppaal} is a tool for validation and verification using graphical simulation and automatic
model-checking of real-time systems. It is based on timed automata, that is finite state machine extended with clocks which are the way to handle time in UPPAAL.

\paragraph{SPIN and ProMeLa\cite{spinprom}}
SPIN is an automated model checker which supports parallel system verification, it is used for analyzing the logical consistency of concurrent systems. The system is described in a modeling language named PROtocol MEta LAnguage (ProMeLa) which provides a vehicle for making abstractions of distributed systems. The intended use of spin is to validate the fractions of process behavior. A complete validation is therefore typically performed in a series of steps, with the construction of increasingly detailed ProMeLa models.



\section{Modeling and verifying using generalized contracts}
\label{section:Methodology}
An issue to be solved for heterogeneous systems is that involved components should conform to an assembly pattern. For the sake of simplicity of the assembly, we adopt the well-researched concept of contract which is therefore extended for the purpose of heterogeneity. Moreover, for a given system we will assume a well \textit{agreed-upon facets} such as data, functionality, time, security, etc.  

\subsection{Definitions}
We extend the traditional A-G contract with the purpose of mastering the modeling and verification of complex and heterogeneous systems.

\begin{definition}[Generalized contract]\label{GC}
	A \textit{generalized contract} is a multi-faceted Assume-Guarantee contract. It is an extension of contract, structured on the one hand with its assume and guarantee parts, and structured on the other hand according to different clearly identified and agreed-upon facets (data, functionality, time, security, quality, etc.) in its assume or guarantee. 
\end{definition}

These facets will be layered to facilitate properties analysis.
In addition, every layer will have a priority. Therefore, an analysis of a facet may be done prior to another facet. 


\begin{definition}[Well-structured component]\label{wsc}
	A well-structured or normalised component is a component equipped with a generalized contract. 
\end{definition}

Therefore, normalising a component $C_i$ consists in transforming $C_i$ into a component equipped with a generalized contract.
A multi-faceted A-G contract can be expressed in PSL for instance.

\subsection{Outline of the proposed method}
\begin{figure*}[!ht]
	\centering
	\includegraphics[scale=0.4]{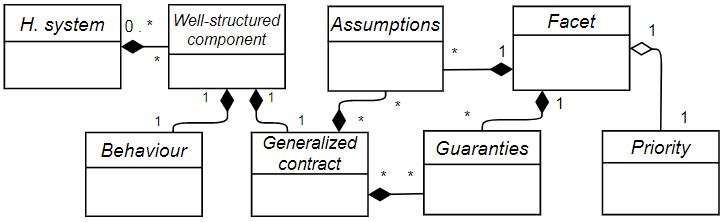}
	\caption{Meta-model of a system with well-structured components}
	\label{figure:meta_model}
\end{figure*}

The working hypothesis is that a heterogeneous system should be an assembly of \textit{well-structured components} (see Def. \ref{wsc}); this is summarized in Figure \ref{figure:meta_model} with the block definition diagrams of SysML \footnote{System modeling Language \cite{sysml}}. Each component has a generalized contract and a behaviour expressed with \textit{labelled transition system} (LTS).  

Starting from requirements and global properties to be fulfilled, the method that we propose (\minarets) consists,  given a set of appropriately selected or constructed elementary components, to normalize these input components prior to their composition, to build a global system, and finally to analyse the global system with respect to the required properties.

For this purpose there are many issues to be solved:\\
\textit{i)} Elementary components are from various languages and cover different facets, a pragmatic mean of composition is required. We choose to adopt a wide purpose expressive language as interface or contract language. Each component will be manipulated through its derived interface (i.e. its \textit{generalized contract} (see Def. \ref{GC})) written in an appropriate language.\\ 
\textit{ii)} Global properties are heterogeneous; they should be clearly expressed, integrated and analysed; we choose to decompose them according to the identified agreed-upon facets and to structure and spread them along the analysis of composed components. The properties will be expressed with a wide purpose expressive language (such as PSL).\\
\textit{iii)} Composition of elementary components should preserve their local requirements and should also be weakened or strengthened with respect to global-level properties. For instance some facets required by an elementary component could be unnecessary for a given global assembly, or some facets required at a global assembly may be strengthen at a component level. \\
\textit{iv)} Global properties require formal analysis tools which are heterogeneous. We choose to separate the concerns, to target various tools and try to ensure the global consistency.\\
\textit{v)} Behaviours of components should be composable.\\

The \minarets method integrates practically the solutions to these issues, as we will present in the sequel. 
We adopt a correct-by-construction approach for the assembly of components. Therefore local compositions should preserve required local properties.
In the same way global properties may impact local components; therefore, global properties are decomposed and propagated through the used components when necessary.


\subsection{Working flow of the \minarets method}
We describe in the following the different steps of our method where the modeling and verification steps are interrelated.


\subsection*{Modeling and verification steps}
\begin{figure*}[ht]
	\begin{center}
		\includegraphics[scale=0.51]{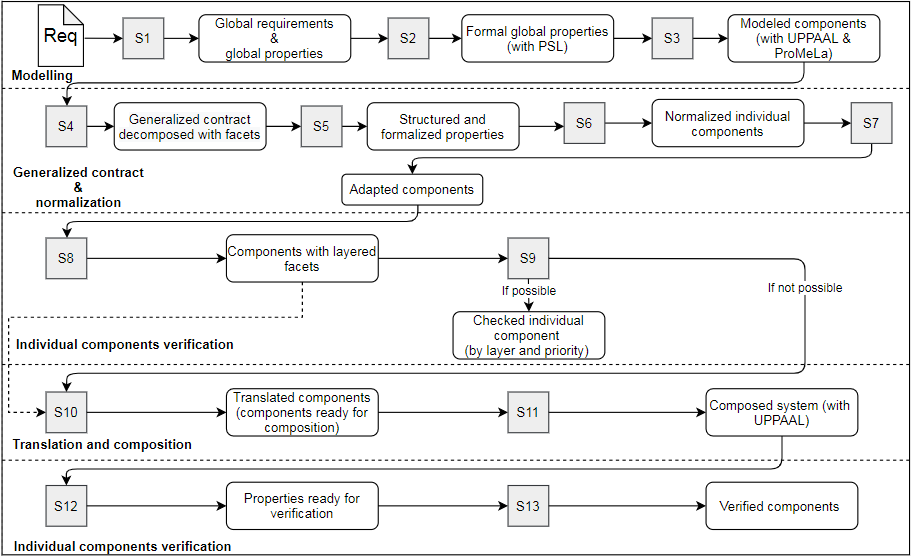}
		\caption{\minarets summarized graph}
		\label{figure:summarized_steps}
	\end{center}
\end{figure*}

\noindent
\textbf{Step 1 \textit{(modeling M$_{1}$)}.} From the informal requirements of the system to build, express the global functional requirements and global properties. They could be expressed with any desired language, even natural one. The only goal here is to \textit{clearly} state the requirements of the given system. State the facets of interaction. \\
\noindent
\textbf{Step 2 \textit{(modeling M$_{2}$)}.} Formalise the required global properties with appropriate expressive formal specification languages (such as PSL). Additionally, state the assume properties of the system. \\
\textbf{Step 3 \textit{(modeling M$_{3}$)}.} Model using the appropriate languages, the individual components or select suitable ones from existing libraries. Accordingly, we will have various heterogeneous components.\\
\textbf{Step 4 \textit{(modeling M$_{4}$)}.} Decompose the global properties with respect to the agreed-upon facets that we considered; this results in a \textit{generalized contract} (the guarantee part) (see Def. \ref{GC}); this will emphasize the concern to be dealt with later on.\\
\textbf{Step 5 \textit{(modeling M$_{5}$)}.} Express the structured properties using the expressive formal specification language.\\
\textbf{Step 6 \textit{(modeling M$_{6}$)}.} Normalize the individual components (that means to equip them with a \textit{generalized contract} (see Def. \ref{wsc}) corresponding to their facets). Some facets may be missing for given components.\\
\textbf{Step 7 \textit{(modeling M$_{7}$)}.} According to the result of \textit{Step 5 (modeling M$_{5}$)}, add a facet of the global property to a component, or ignore some of its facets, if necessary. Consequently the component should be rechecked.\\
\textbf{Step 8 \textit{(modeling M$_{8}$)}.} Order (or attribute a priority) to each facet for global verification purpose. Therefore we have \textit{layers} of facets; for instance, $Layer1\_Data$, $Layer2\_Time$, $Layer3\_Security$...etc. The verification by layer allows one to verify the contracts by order; from a very important layer (primary) to a less important layer (secondary). If the behaviour of our system does not satisfy a primary layer of contract, then, it is not necessary to continue the verification with the other layers of contracts as well as properties. Prioritising can be done among the property of a layer.  \\
\textbf{Step 9 \textit{(verification V$_{1}$)}.} Check the good functioning of each normalized individual component if tools exist for that and if the required data are available.\\
\textbf{Step 10 \textit{(modeling M$_{9}$)}.} If the checking of the normalized individual components cannot be carried out, due to lack of tools or the component need to be assembled to be verified, then a translation of the component to a target language is necessary; after that, the component is ready for composition.\\
\textbf{Step 11 \textit{(modeling M$_{10}$)}.} Compose the normalized components if they are composable. As we focus on behaviours expressed with LTS, the composition results in parallel composition with the appropriate formalism.\\ 
\textbf{Step 12 \textit{(verification V$_{2}$)}.} Translate for each individual component, its \textit{generalized contract} from the wide purpose language into the language of the targeted model-checker or simulator that will be used.\\
\textbf{Step 13 \textit{(verification V$_{3}$)}.} Apply the verification by layer with respect to the \textit{specified order}. Properties of the same layer are checked together.\\
\\
\noindent
Finally, we present in Figure \ref{figure:summarized_steps} a summarized graph of our \minarets method.

\subsection{Implementation of the method}
Concerning the wide purpose language to express the heterogeneous formal properties we have two choices: \\
a) To develop a new language that would allow one to specify the various properties we wanted to use in a formal, simple and unambiguous way; to be able to write a global properties, as well as to allow one to use it as an input for formal verification by model-checking or simulation; finaly we will  need to develop a tool that would support the language.\\  
b) To choose a language and a tool that respond to our needs and that already exist in the community.

The last choice is made. For efficiency we choose the PSL standard \cite{PSL} that suits us in our research as well as the Active-HDL tool that supports it. Therefore we consider some experimentations with tools-based formalisms such us PSL, ProMeLa/SPIN, UPPAAL.   

For the verification of the properties of components and systems, we have two choices, if an efficient tool exists to verify the built or selected component, the verification will be done with it; otherwise, a translation of the component and its properties to the targeted tool is needed. 
Currently the targeted tools are for example the model checkers UPPAAL, SPIN.

\section{Case study 1}
\label{section:study-case}
\begin{figure}[H]
	\begin{center}
		\includegraphics[scale=0.39]{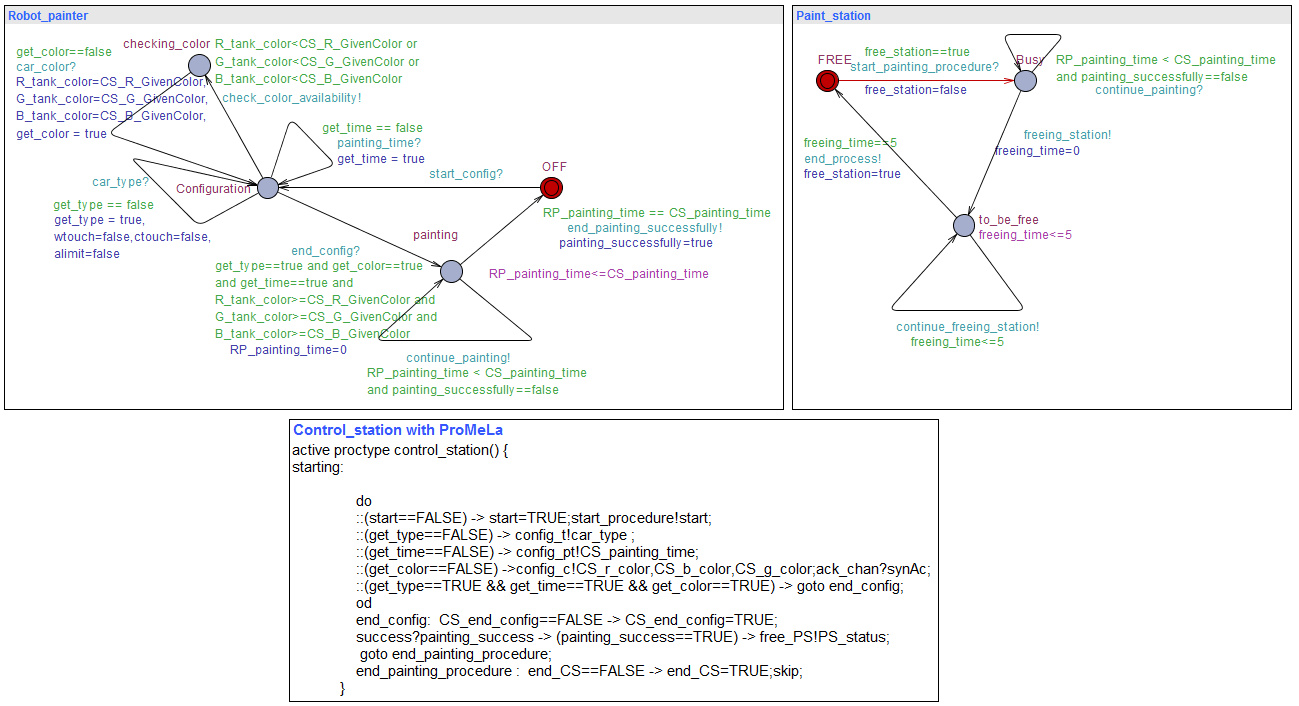}  
		\caption{ PS and RP models with UPPAAL and CS with ProMeLa }
		\label{figure:modeles_des_composants}
	\end{center}
\end{figure}

\begin{figure}[H]
	\begin{center}
		\includegraphics[height=0.2\textheight,width=0.9\linewidth]{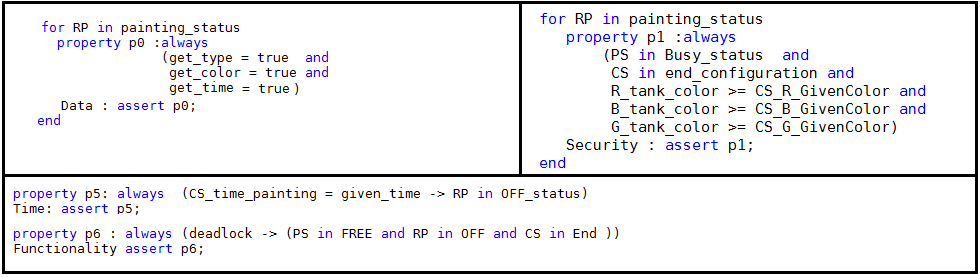}  
		\caption{A part of faceted and formalized global property with \textit{PSL} and \textit{Aldec Active-HDL }} 
		\label{figure:glob_prop_psl}
	\end{center}
\end{figure}

\begin{figure}[H]
 	\begin{center}
 		\includegraphics[scale=0.75]{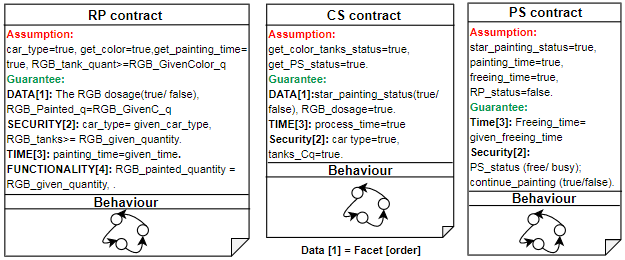}  
 		\caption{Normalized components with the prioritised facets}
 		\label{figure:normalized_components}
 	\end{center}
 \end{figure}

\begin{figure}[!htb]
		\includegraphics[scale=0.36]{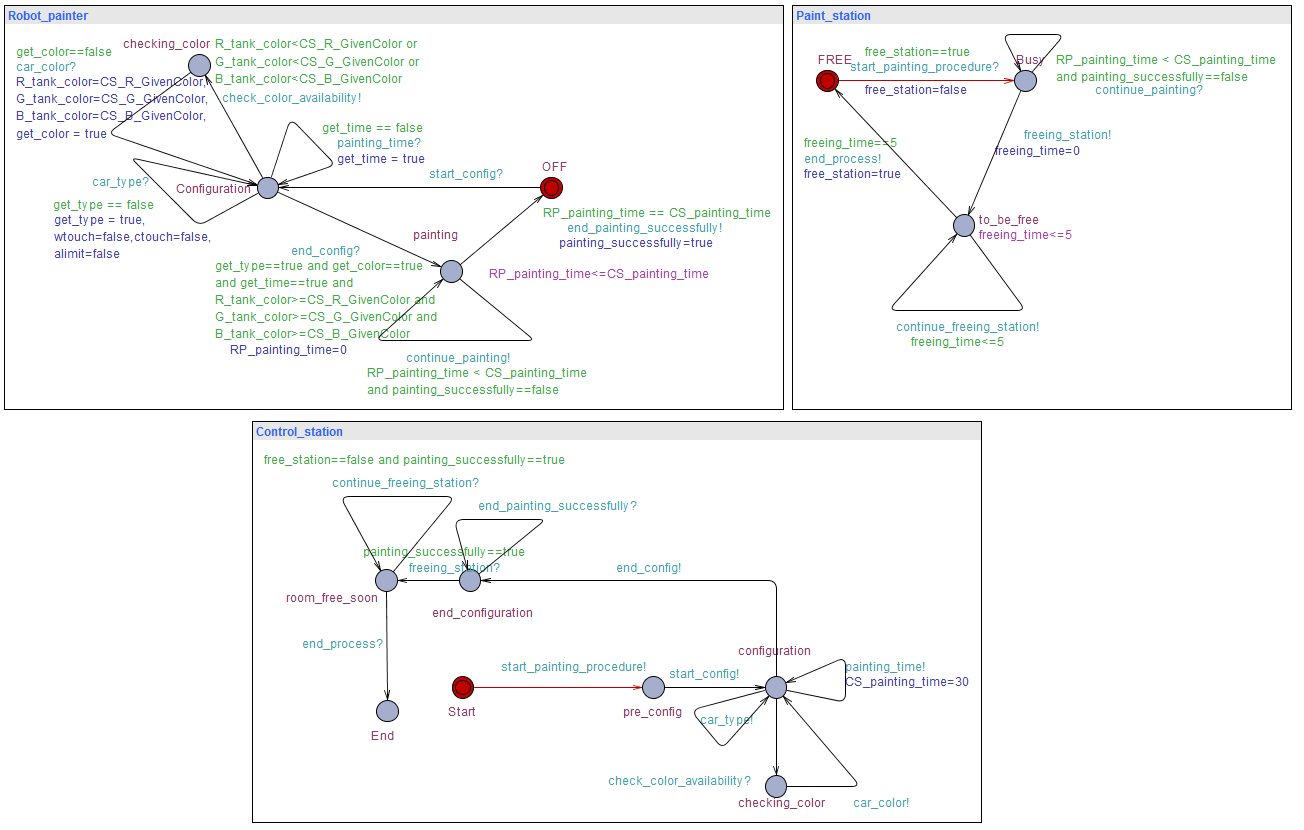}  
		\caption{The composed system after the component translation (in UPPAAL)}
		\label{figure:composed_system}
\end{figure}

\begin{figure}[H]
	\begin{center}
		\includegraphics[scale=0.5]{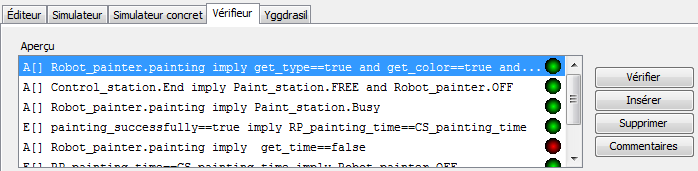}  
		\caption{Status of properties verification with UPPAAL}
		\label{figure:verification_UPPAAL}
	\end{center}
\end{figure}

\paragraph{} 
To illustrate the use of the proposed method, we consider a case study of an automotive industry (painting workshop). This workshop is composed of three principal components; the control station (CS) manages the paint station (PS) and the automatic robot painter (RP); these components cooperate with each other to achieve the painting process correctly; achieve the painting process correctly means to paint the cars with the desired color, in time, without any damages and without wasting color.\\
For that, the components CS, PS and RP send the necessary informations to each other and receive other informations at the same time. A violation of one or more information can cause a significant material damage and human also. An example of these informations (car type, PS status, RP and tanks status); If RP doesn't respect the "car type" it means that RP doesn't know the dimension of the car, it is not well configured, so, RP could damage the car or hit persons who are in the PS; these informations are  sensitive; we have various constraints for different status like the status of color\_tanks, if RP works without color in the tanks, the robot could be damaged...etc.\\
\\
Now we follow step by step the \minarets method to model and verify this system.\\
\\
\noindent
 Applying \textbf{Step 1 \textit{(modeling M$_{1}$)},} we obtain the following informal global requirements and global properties:\\
\textbf{Global requirements:} cars informations (type, color with RGB quantity, painting time); sufficient RGB colors in the tank; freeing time must be defined for each car.\\
\textbf{Global expected properties:} the system respects the correct RGB dosage; there is no loss of color; the painting time should be equal to the given time; the painting is done without damages (respect of car dimensions, it is known from the car type); freeing time must be equal to the given freeing time for each car; stop and notify when there is no sufficient color; painting station status must be free before use, and busy when a car is inside; the painting starts after the end of configuration, and it finishes when the painting time equal to given painting time.  \\
 After \textbf{Step 2 \textit{(modeling M$_{2}$)},} we obtain the following formalized global properties.\\
\textbf{Global requirements:\\}
\texttt{
\hspace*{0.8cm} get\_type = true;\ get\_color\ =\ true;\ get\_painting\_time\ =\ true; \\
\hspace*{0.8cm} R\_tank\_quantity\ >=\ R\_GivenColor\_quantity,\\
\hspace*{0.8cm} G\_tank\_quantity\ >=\ G\_GivenColor\_quantity,\\
\hspace*{0.8cm} B\_tank\_quantity\ >=\ B\_GivenColor\_quantity;\\
\hspace*{0.8cm}\ get\_freeing\_time\ =\ true;\\}
\textbf{Global properties:\\}\texttt{
\hspace*{0.8cm} R\_color\_PaintedQuantity\ =\ R\_GivenColor\_quantity, \\
\hspace*{0.8cm} G\_color\_PaintedQuantity\ =\ G\_GivenColor\_quantity,\\
\hspace*{0.8cm} B\_color\_PaintedQuantity\ =\ B\_GivenColor\_quantity; \\
\hspace*{0.8cm} painting\_time\ =\ GivenPainting\_time;\ car\_type\ =\ given\_type;\\
\hspace*{0.8cm}\ freeing\_time\ =\ given\_freeing\_time;\\
\hspace*{0.8cm}/*each color is controlled separately*/\\
\hspace*{0.8cm} if\ (RGB\_tank\_quantity\ <=\ RGB\_GivenColor\_quantity)\\ 
\hspace*{0.8cm} Then\ (stop\_process\ and\ warning\_message);  \\
\hspace*{0.8cm} if\ (car)\ in\ painting\ Then \ PaintingStation\_status\ =\ Busy; \\
\hspace*{0.8cm} and\ end\_configuration\ =\ true;\\
\hspace*{0.8cm} else\ PaintingStation\_status\ =\ free;\\
\hspace*{0.8cm} painting\_time = GivenPainting\_time\ imply\ painting\ = \ finished;\\}
 \textbf{Step 3 \textit{(modeling M$_{3}$)}.} Here we choose the UPPAAL model checker and the SPIN tool with its ProMeLa language to model the components RP, CS and PS. Figure \ref{figure:modeles_des_composants} shows the models under both environments.\\
\noindent
Applying \textbf{Step 4 \textit{(modeling M$_{4}$)},} we obtain the generalized contract decomposed with the facets (Data, functionality, time, security). The Figure \ref{figure:glob_prop_psl} shows a part of the faceted and formalized properties. (The guarantee part).\\
 Applying \textbf{Step 5 \textit{(modeling M$_{5}$)},} we obtain the structured and formalized properties with PSL language as depicted in Figure \ref{figure:glob_prop_psl}.\\
\noindent
 Applying the \textbf{Step 6 \textit{(modeling M$_{6}$)}} for normalization; we integrate the assumptions and guarantees of each individual component. Figure \ref{figure:normalized_components} shows the normalized individual components. \\
CS requires tanks status and PS status, on the other hand, CS guarantees the start painting status, and it sends car type, RGB dosage, painting time to RP and must feel the tanks of colors if needed. With the received data, RP in turn guarantees the correct painting of cars in time, without scratching them, without wasting color, it doesn't work without sufficient color in tanks. PS also require start painting status, process time and RP status, and in return, it guarantees the painting of one car at the same time by managing his status busy/free, and the freeing time. \\
 Applying \textbf{Step 7 \textit{(modeling M$_{7}$)},} in the current case it is not necessary to add or ignore a facet. We have suitable components ready for verification and composition. see Figure \ref{figure:normalized_components}.\\
\noindent Applying \textbf{Step 8 \textit{(modeling M$_{8}$)},} we attribute the following priorities to each facet (data=1, security=2, time=3 functionality=4; were number 1 is the highest priority). We obtain ordered layers with respect to facets and properties as well. the order of layer is mentioned in Figure \ref{figure:normalized_components}.\\
\noindent
\textbf{Step 9 \textit{(verification V$_{1}$)},} in our case  as we use ProMeLa we have an adequate tool (SPIN) but, the only component RP modelled with ProMeLa language cannot be verified without composition with its environment.\\
 Applying \textbf{Step 10 \textit{(modeling M$_{9}$)},} we translate the ProMeLa component RP to UPPAAL using the algorithm \ref{alg1}, we obtain a component RP ready for composition (see Figure \ref{figure:composed_system}).\\
 Applying \textbf{Step 11 \textit{(modeling M$_{10}$)},} we compose the translated component RP with the other components CS and PS with the UPPAAL tool. We obtain the composed system depicted in Figure \ref{figure:composed_system}.\\
\noindent Applying \textbf{Step 12 \textit{(verification V$_{2}$)},} we translate the generalized contract from \textit{PSL} into the language of UPPAAL. We obtain a suitable properties ready for verification. The following property is an extract of the translated ones.\\
\texttt{
\hspace*{1.5cm} {{A[\ ] \ Robot\_painter.painting\ imply\ get\_type\ ==\ true\ }}  \\
\hspace*{2cm}{{and\ get\_color\ ==\ true\ and\ get\_time\ ==\ true}}\footnote{were "A [ ] Prop" denotes the "always property".}  \\}
Applying \textbf{Step 13 \textit{(verification V$_{3}$)},} we verify each component by layer: data, security, time, functionality; also, the primary properties before the secondary. At the end of this step, we obtain the verified components. Figure \ref{figure:verification_UPPAAL} shows the verification status of the translated properties.   \\

\section*{Case study 2}
\begin{figure}[H]
	\begin{center}
		\includegraphics[scale=0.55]{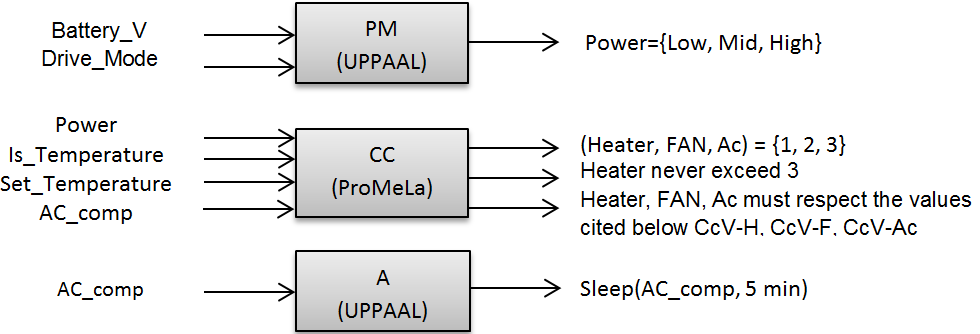}  
		\caption{The selected components PM and A modelled with UPPAAL and CC with ProMeLa}
		\label{figure:selected_c}
	\end{center}
\end{figure}

The functionality of the ClimateControl component is to set the correct values for the Heater, Fan, and ACCompressor, which are responsible to interact with corresponding hardware components. The ClimateControl sets these values depending on its inputs: the current temperature provided by TemperatureIsSensor, the set temperature provided by TemperatureSetSensor, the information about the air condition button state (ACOnOffMode), the currently available energy provided by the PowerManagement component. Depending on the battery voltage (BatteryUSensor) and the set drive mode (DriveModeSensor), e. g. Battery and eco or sport mode, The PowerManagement component returns either HIGH, MID, or LOW, which are used as different levels of energy consumption. Finally the Airing component is called to stop the Air-condition compressor for 5 min each 30 minutes \cite{kugele}.\\
\textbf{Step 1 (modeling M$_{1}$)}, We express the informal global requirements and global properties.
They could be expressed with any desired language, even natural one. The only
goal here is to clearly state the requirements of the given system.\\
\noindent \textbf{Global requirements:} to work properly the Climate controller requires: \\
The battery voltage; The drive mode; The actual temperature; The set temperature and air-condition status.\\
\noindent\textbf{Global expected properties:}\\
PmV: the Power manager $\in$ (LOW, MID, HIGH).\\
HV:  the Heater values $\in$ (1, 2, 3).\\
FV:  the Fan values $\in$ (1, 2, 3) .\\
AcV: the Air condition compressor values $\in$ (1, 2, 3) . \\
CcV-H: the Climate controller must give the correct value to the Heater.\\
CcV-F: the Climate controller must give the correct value to the Fan.\\
CcV-Ac: the Climate controller must give the correct value to the Air-condition.\\
Hmax: the heater never exceed the values 3. \\
AcS: the air-condition compressor stops working 5 min each 30 min.\\
{\scriptsize
	PmV: the Power manager values.\\
	HV: the heater values interval.\\
	FV: the FAN values interval.\\
	AcV: the Air-condition values interval.\\
	CcV-H: the Climate controller value for Heater.\\
	CcV-F: the Climate controller value for the Fan.\\
	CcV-Ac: the Climate controller value for Air condition. \\
	Hmax: the maximum value for the Heater.\\
	AcS: the air-condition Sleep.\\}
\textbf{Step2 (modeling M$_{2}$)}, We formalise the required global properties with appropriate
expressive formal specification languages (PSL), we obtain the following
formalized global properties. \\
\textbf{Global requirements:}\\
\texttt{\hspace*{0.8cm}Battery\_V\ <=12\ and\ Battery\_V >=\ 0;   // Battery voltage  \\
\hspace*{0.8cm}Drive\_Mode\ <=5\ and\ Drive\_Mode\ >=\ 0;     // Drive mode \\
\hspace*{0.8cm}IS\_Temperature\ <= 50\ and\ IS\_Temperature>=\ -30;   // Actual temperature \\
\hspace*{0.8cm}Set\_Temperature\ <=\ and\ Set\_Temperature\ >=\ 16;    // Set temperature \\
\hspace*{0.8cm}AC\_comp\ =\ On\ or\ AC\_comp\ =\ OFF;    // Air-condition compressor \\}
\textbf{Global properties:}\\
\texttt{
\hspace*{0.8cm}Power in \{LOW,\ MID,\ HIGH\};           // PmV \\
\hspace*{0.8cm}Heater in \{1,\ 2,\ 3\};	     // HV \\
\hspace*{0.8cm}FAN in \{1,\ 2,\ 3\};			     // FV\\
\hspace*{0.8cm}Ac\ in \{1,\ 2,\ 3\};		                  // AcV	\\
Heater, FAN, AC\_comp\_v must be correct, it means:  \\  
/* CcV-H */\\
\hspace*{0.8cm}AC\_comp=\ OFF\ and\ Power=LOW\\ \hspace*{0.8cm} and\ (IS-set\ temperature)<0->Heater=1;\\
\hspace*{0.8cm}AC\_comp=\ OFF\ and\ Power=MID\ \\\hspace*{0.8cm} and\ (IS-set\ temperature)<0->Heater=2;\\
\hspace*{0.8cm}AC\_comp=\ OFF\ and\ Power=HIGH\\\hspace*{0.8cm} and\ (IS-set\ temperature)<0->Heater=3;\\
/* CcV-F   */\\
\hspace*{0.8cm}AC\_comp=\ ON\ and\ Power\ =\ LOW\\\hspace*{0.8cm} and\ (IS-set\ temperature)>=0\ ->FAN=1;\\
\hspace*{0.8cm}AC\_comp=OFF\ and\ Power=HIGH\ \\\hspace*{0.8cm}and\ (IS-set\ temperature)>=0\ ->FAN=3;\\
(……We limit the cases……)\\
/* CcV-Ac  */\\
\hspace*{0.8cm}AC\_comp=\ ON\ and\ Power\ =\ LOW\\\hspace*{0.8cm} and\ (IS-set\ temperature)\ >=\ 0\ ->\ Ac\ =\ 1;\\
\hspace*{0.8cm}AC\_comp=\ ON\ and\ Power\ =\ MID\\\hspace*{0.8cm} and\ (IS-set\ temperature)\ >=\ 0\ ->\ Ac\ =\ 2;\\
\hspace*{0.8cm}AC\_comp=\ ON\ and\ Power\ =\ HIGH\\\hspace*{0.8cm} and\ (IS-set\ temperature)\ >=\ 0\ ->\ Ac\ =\ 3;\\
/* AcS */\\
\hspace*{0.8cm}If\ (AC\_comp\_working\ =\ 30\ min)\\\hspace*{0.8cm} then\ sleep(AC\_comp,\ 5min);\  end    \\
/* Hmax */\\
\hspace*{0.8cm}Heater\ never\ exceed 3;     \\}
\textbf{Step 3 (modeling M$_{3}$)}. We select the components PowerManager (PM), ClimateControl (CC) and Airing (A) from our software library. (See Figure \ref{figure:selected_c}).\\
\noindent\textbf{Step 4 (modeling M$_{4}$)}, We decompose the global properties with respect to the agreedupon-
facets that we considered (Data, functionality, time, security); we obtain the generalized contract decomposed with the facets.\\
\textbf{The guarantee part:}\\
PmV:\texttt{ Power\ in \{LOW,\ MID,\ HIGH\};  //\textbf{DATA}\\}
HV:\texttt{ Heater\ in \{1,\ 2,\ 3\};          //   \textbf{DATA}\\    }  
FV: \texttt{ FAN\ in\ \{1,\ 2,\ 3\};		       //\textbf{DATA}\\}
AcV: \texttt{ AC in \{1,\ 2,\ 3\};         		     //  \textbf{DATA}\\}
Heater, FAN, Ac must respect the correct values: \\
\texttt{(see properties CcV-H, CcV-F, CcV-Ac)    // \textbf{FUNCTIONALITY}\\}
AcS:\texttt{ If\  (AC\_comp\_working\ =\ 30\ min)\    
then\ sleep(AC\_comp,\ 5min);\ end  // \textbf{TIME}\\}
Hmax:\texttt{ Heater\ never\ exceed\ 3               //  \textbf{SECURITY}\\}

\begin{figure}[H]
	\begin{center}
		\includegraphics[scale=0.6]{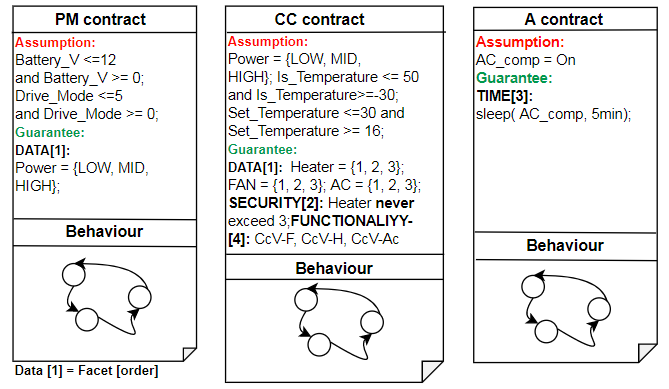}  
		\caption{Normalized components}
		\label{figure:normalized_c}
	\end{center}
\end{figure}

\begin{figure}[H]
	\begin{center}
		\includegraphics[scale=0.4]{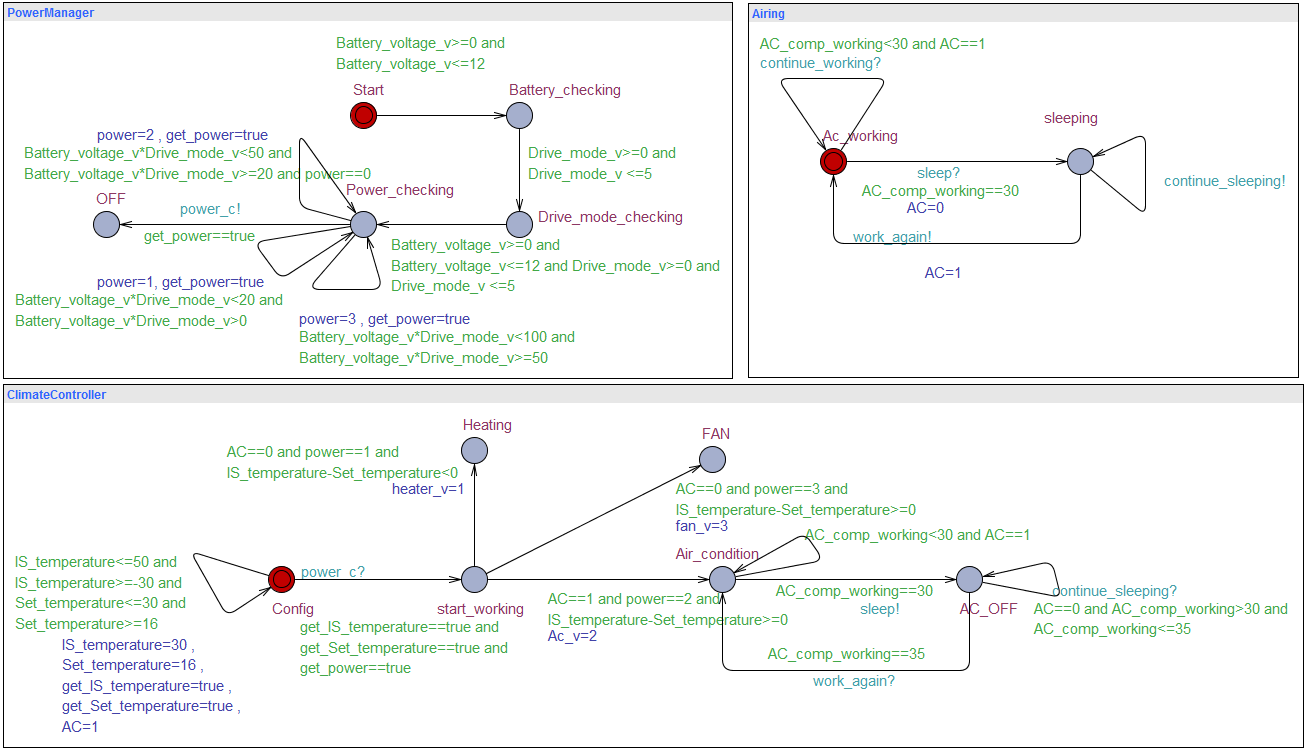}  
		\caption{The composed system after the component translation (in UPPAAL) }
		\label{figure:composed_case2}
	\end{center}
\end{figure}

\begin{figure}[H]
	\begin{center}
		\includegraphics[scale=0.4]{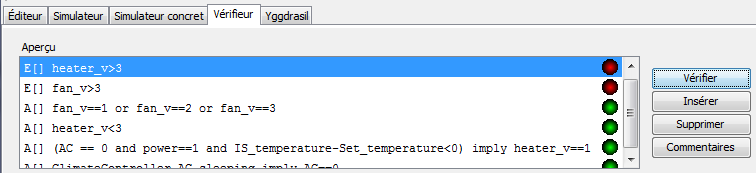}  
		\caption{Status of properties verification with UPPAAL}
		\label{figure:verif_2}
	\end{center}
\end{figure}

\noindent \textbf{Step 5 (modeling M$_{5}$)}, we structure and formalize the properties with PSL language. We obtain the following structured and formalized properties.\\
\texttt{
\noindent\hspace*{0.8cm}\textcolor{blue}{Property} PmV: \textcolor{blue}{always} Power \textcolor{blue}{in} \{LOW, MID, HIGH\};\\
	\hspace*{0.8cm}\textcolor{blue}{Property} HV: \textcolor{blue}{always} Heater \textcolor{blue}{in} \{1, 2, 3\}; \\
	\hspace*{0.8cm}\textcolor{blue}{Property} FV: \textcolor{blue}{always} FAN \textcolor{blue}{in} \{1, 2, 3\}; \\
	\hspace*{0.8cm}\textcolor{blue}{Property} AcV: \textcolor{blue}{always} Ac \textcolor{blue}{in} \{1, 2, 3\};\\
	\hspace*{0.8cm}Data: \textcolor{blue}{assert} PmV \textcolor{blue}{and} HV \textcolor{blue}{and} FV \textcolor{blue}{and} AcV;\\
}Heater, FAN, AC\_comp must respect the correct values:    \\ \texttt{
\hspace*{0.8cm}\textcolor{blue}{Property} CcV-Ac: \\
	\hspace*{0.8cm}\textcolor{blue}{if} ( AC\_comp = ON \textcolor{blue}{and} Power = Low \textcolor{blue}{and}\\
	\hspace*{0.8cm}IS\_Temperature - Set\_Temperature >=0) \textcolor{blue}{then} AC = 1;\\
	\hspace*{0.8cm}\textcolor{blue}{else if} ( AC\_comp = ON \textcolor{blue}{and} Power = Mid \textcolor{blue}{and}\\
	\hspace*{0.8cm} IS\_Temperature - Set\_Temperature >=0) \textcolor{blue}{then} AC = 2;\\
	\hspace*{0.8cm}\textcolor{blue}{else if} ( AC\_comp = ON \textcolor{blue}{and} Power = High \textcolor{blue}{and}\\
	\hspace*{0.8cm} IS\_Temperature - Set\_Temperature >=0) \textcolor{blue}{then} AC = 3; \textcolor{blue}{end} \\
	\hspace*{0.8cm}Functionality : \textcolor{blue}{assert} CcV-Ac;\\
	\hspace*{0.8cm}/* (…CcV-F, CcV-H…)*/\\
	\hspace*{0.8cm}\textcolor{blue}{Property} AcS : \textcolor{blue}{if}  (AC\_comp\_working =  30 min)\\
	\hspace*{0.8cm}  \textcolor{blue}{then} sleep (AC\_comp , 5min);  \textcolor{blue}{end}\\
	\hspace*{0.8cm}Time : \textcolor{blue}{assert} AcS;\\
	\hspace*{0.8cm}\textcolor{blue}{Property} Hmax : \textcolor{blue}{never} Heater > 3;\\
	\hspace*{0.8cm}Security : \textcolor{blue}{assert} Hmax;\\}
\textbf{Step 6 (modeling M$_{6}$)} Normalizing the individual components, we integrate the assumptions
and guarantees of each individual component.(See Figure \ref{figure:normalized_c})\\
\noindent\textbf{Step 7 (modeling M$_{7}$).} According to the result of Step 4 (modeling M4), we may add a
facet of the global property to a component, or ignore some of its facets, if necessary.
In the current case study it is not necessary to add or ignore a facet. We have suitable components PM, CC and A. (See Figure \ref{figure:normalized_c}).\\ 
\textbf{Step 8 (modeling M$_{8}$),} we attribute the following priorities to each facet
(data=1, security=2, time=3 functionality=4; were number 1 is the highest priority).We
obtain an ordered layers with respect to facets and properties as well.(See Figure \ref{figure:normalized_c})\\
\textbf{Step 9 (verification V$ _{1} $):} in our case as we use ProMeLa we have an adequate tool
(SPIN) but, the component ClimateControle only modelled with ProMeLa language cannot be verified
without composition with its environment. (it needs composition with other components). \\
Applying\textbf{ Step 10 (modeling M$ _{9} $),} we translate the ProMeLa component ClimateControl to UPPAAL
using the algorithm 1, we obtain a component ClimateControl ready for composition. (the component ProMeLa code is in Appendix).\\
Applying \textbf{Step 11 (modeling M$ _{10} $),} we compose the translated component ClimateControl with the other components PowerManager and Airing with the UPPAAL tool. We obtain the composed system depicted in Figure \ref{figure:composed_case2}. (notice that because of combinatory explosion we limited the cases for CC component).\\
\noindent Applying \textbf{Step 12 (verification V$ _{2} $)}, we translate the generalized contract from PSL
into the language of UPPAAL. We obtain suitable properties ready for verification. See figure
The following property is an extract of the translated ones.\\
\texttt{
\hspace*{2.5cm}A[\ ](AC\_comp==ON\ and\ Power==Low\ and \\
\hspace*{2.5cm}\ IS\_Temperature-Set\_Temperature>=0\\
\hspace*{2.5cm}\ imply\ FAN == 1\ and\ AC\_comp\_v == 1)\\}
Applying \textbf{Step 13 (verification V$ _{3} $)}, we verify each component by layer: data, security,
time, functionality; also, the primary properties before the secondary (the level with highest priority is verified first). At the end of this step, we obtain the verified system. Figure \ref{figure:verif_2} shows the verification status of the translated properties.

\paragraph{\textbf{Assessment}}

This experimentation was conducted with the sake of improvement of our method in mind.
We have considerably detailed the steps of the method when thinking thoroughly about its applicability through the case study.
Despite the success of applying in preliminary trivial exercises and on this case study, and the reproducibility of the steps, it appears that more tools assistance is needed to guide the users.
The experimentations even if not yet scalable to industrial cases, gave the opportunity to tune the method steps and design some translators to help in modeling and verification.
We are aware of the impact of treated facets on interactions between various tools;
but more lessons from various case studies will help to promote good practices through tools. Our research report \cite{minarets} contains other detailed experimention examples. We already planned some investigations on other combinations of facets and tools.

\section{Related work}
\label{section:Related_work}

There are several works that contribute to the heterogeneity issues, they propose different methods and techniques.
Interface theory \cite{henzinger2001} and theories of contracts \cite{Benveniste2007, benveniste2012, Benveniste2014, Benveniste2018}, are based on the contracts of components and the informations given by their interfaces; they use intensively contracts as well as on behaviour; they propose techniques and methods to promote concurrent development, such as the mathematical foundations of contract model presented in \cite{Benveniste2007} which are developed in the SPEED \footnote{SPEculative and Exploratory Design in systems engineering http://www.speeds.eu.com}  project context that is concerned with the development of an illustrative methodology with tools to promote speculative design.

In the Ptolemy project, \cite{Lee2010} proposes an approach of interaction between heterogeneous components based on models of calculation (MOC); here the heterogeneity is linked to different models of calculation. These MOCs have a concrete semantics which is transformed into abstract semantics with three aspects: execution control, communication and time models \cite{Lee2010}.
Ptolemy proposes a solution of interaction between several actor-oriented models, the interaction acts as a global scheduler; Ptolemy uses abstract semantics, in fact, they do not use actor-oriented models but rather finite state machines. From our point of view, this composition method is heavy, general and constrains the use of contracts, especially when we deal with the growing small aperture nets of processes \cite{Att2017}. In our \minarets method we aim to exploit more advantages of contracts in a simpler and explicit way to model and verify the assembly of heterogeneous components.

In \cite{Mes2010} our team members have contributed to the improvement of incremental component composition manipulation and verification of the interaction between heterogeneous components. Here the heterogeneity is related to granularity and not to languages. They were based on the communication that goes down or up in the hierarchy of a composite; it means: the interaction between the internal components of a composite.
They studied the layering of the different levels of structures of a contract (service, component, assembly, composition) on the different levels of requirements (Static, Architectural, Functional, behavioural, quality of service), with the aim of mastering the incremental assembly of components and verifying multi-aspect properties; from where, the levels of interoperability between the different languages of each component have been defined by the multi-layer contracts.

Our modeling and verification method is based on our team results \cite{Att2017} for heterogeneous components composition. They deal with the formal composition of heterogeneous components. They are based on the dynamic behaviour of the components; moreover, they have chosen the labelled transition systems as common semantic domain; the proposed method is focused on the manipulation of the channels, which are the communication mechanism between the assembled models based on an algebra of operators. they illustrated their approach by aZIZA tool \footnote{https://aziza.ls2n.fr/ \cite{Att2017}}. 

As a part of B.I.P project, 
\cite{Sif2014} proposes a technique with three layers: "\textit{Behavior, Interactions, Priorities}" (B,I,P). The first layer (B) describes the dynamic behaviour of the component; it consists of a set of labelled transition systems (LTS). The second interaction layer (I) is responsible for the coordination of component execution; and the third priority layer (P) provides a mechanism for restricting the overall behaviour of the lower layers with a filtering of possible interactions\cite{Sif2014}.
\cite{Sif2014} proposes a low level solution that deals with the interaction between components; it focuses on the composition of the system with the different interaction semantics, however, unlike \minarets they didn't deal with the heterogeneous components modeling but they deal with the property analysis and composition of components using B.I.P language.





\section{Conclusion}
\label{section:conclusion}

\noindent We have proposed the \minarets method for complex and heterogeneous systems modeling and analysis; it is based on an extension of the traditional contract, resulting in generalized contracts, which consists in organizing contracts into several facets, depending on the nature of the properties that we are dealing with. \minarets emphasizes the composition of heterogeneous components equipped with generalized contracts. We have shown how one can reduce the complexity of the global modeling and the global analysis of complex and heterogeneous systems.

We illustrated our approach with an example of a paint workshop in the automotive industry domain; it involves different facets (data, functionality, security, time). We have checked the properties concerning the various facets. Concerning verification, generalized contracts are first expressed in PSL, then translated into input languages of UPPAAL/SPIN model checkers.

A direction of future work is not only the scalability but also to study various policies for the composition of contracts; it is important to verify the components composition as we done through the used tools but, as part of our method improvement we need to propose and perform the parallel contracts composition in various formal ways. This will furthermore strength the foundations of our method, and enable the contract management tool construction.


\bibliographystyle{plain}

\bibliography{\repBIBLIO/biblioFile}

\newpage
\begin{appendix}
{\noindent \huge \textbf{Appendix}}
\section{ProMeLa to UPPAAL Translation Algorithm}
\begin{algorithm}[H]
	\caption{Translate ProMeLa to UPPAAL}
	\label{alg1}
	
	\begin{algorithmic}[1]
		\Start
		\State{\read pml-file ;}	
		\State{ $ Components    \gets  find$  all Transition Systems (TS) from processes declarations and  instantiation ;}
		
		\ForAll{$  TS  \in components $ }
		\State \print system declarations and initializing information for TS ;	
		\If {declarations\_type = TIMER\_X}
		\State $ declaration\_type  \gets clock $ and do not increment manualy   ;
		\EndIf
		\ForAll { outgoing transition (goto) in label }{
			\ForAll{ $guard \in transition$ }
			\State \print guard;
			\EndFor
			\ForAll{ $synchronization \in transition$}{
				\State \print synchronization ;}
			\EndFor 
			\ForAll{ $assignment \in transition$}
			\If{Guard = True}
			\State \print code that makes assignments ; 
			\EndIf 
			\EndFor
			
		}
		\EndFor
		\EndFor
		\End
	\end{algorithmic}
	
\end{algorithm}

\section{The ProMeLa component code}
\texttt{
\noindent Active proctype Climate\_controller()\{\\
	confing:\\
	do\\
	::(IS\_temperature<=50 and IS\_temperature>=-30 and\\
	Set\_temperature<=30 and Set\_temperature>=16) -> \\
	IS\_temperature=30 and Set\_temperature=16 and\\
	get\_IS\_temperature=true and get\_Set\_temperature=true and\\
	AC=1; goto config;\\
	:: (get\_IS\_temperature==true and get\_Set\_temperature==true \\
	and get\_power==true) -> power\_c?; goto start\_working;\\
	od\\
	start\_working:\\
	do\\
	/*Cases are limited because of combinatory explosion */\\
	:: (AC==0 and power==1 and IS\_temperature-Set\_temperature<0) -> goto Heating; heater\_v=1;\\
	:: (AC==0 and power==3 and IS\_temperature-Set\_temperature>=0) -> goto FAN; fan\_v=3;\\
	:: (AC==1 and power==2 and IS\_temperature-Set\_temperature>=0) -> goto Air\_condition; Ac\_v=2;\\
	od\\
	Air\_condition:\\
	do\\
	:: (AC\_comp\_working<30 and AC==1) -> goto Air\_condition;\\
	:: (AC\_comp\_working==30) -> sleep!; goto AC\_sleeping;\\
	od\\
	AC\_sleeping:\\
	do\\
	:: (AC==0 and AC\_comp\_working>30 and \\
	AC\_comp\_working<=35) -> continue\_sleeping? ;\\
	:: (AC\_comp\_working==35) -> work\_again?; goto Air\_condition;\\
	od\\
Heating:\\
FAN:\\
\}
}

\end{appendix}

  
\end{document}